\documentclass[twocolumn,prl]{revtex4-1}
\usepackage{hyperref}
\usepackage{amssymb,amsmath}
\usepackage{graphicx}
\usepackage{rotating}
\usepackage{color}

\begin{document}

\title{Soliton--antisoliton pair production in particle collisions}

\author{S.V. Demidov}\email{demidov@ms2.inr.ac.ru}
\affiliation{Institute for Nuclear Research of the Russian Academy of
  Sciences, 60th October Anniversary prospect 7a, Moscow 117312,
  Russia}
\author{D.G. Levkov}\email{levkov@ms2.inr.ac.ru}
\affiliation{Institute for Nuclear Research of the Russian Academy of
  Sciences, 60th October Anniversary prospect 7a, Moscow 117312,
  Russia}

\begin{abstract}
  We propose general semiclassical method for computing the probability
  of soliton--antisoliton pair production in particle collisions.
  The method is illustrated by explicit numerical calculations
  in $(1+1)$--dimensional scalar field model. We find that the
    probability of the process is suppressed by an exponentially small
    factor which is almost constant at high energies.
\end{abstract}

\maketitle

Ever since the discovery of topological solitons, a
question remains~\cite{Drukier,solitons,*solitons1} of whether
soliton--antisoliton (SA) pair can be produced at sizable
probability in collision of two quantum particles. This 
process, which involves a transition from perturbative two--particle state to
a non--perturbative state containing SA pair, eludes treatment by any
of the standard methods. A general 
expectation~\cite{Drukier,unitarity,*unitarity1,*unitarity2} is
that the 
probability of the process is exponentially suppressed in
weak coupling regime,
\begin{equation}
\label{eq:1}
{\cal P}(E) \sim \mathrm{e}^{-F(E)/g^2}\;,
\end{equation}
where $E$ is the total energy,  $g\ll 1$ is the coupling
constant. Indeed, crudely speaking, one can think
of solitons as bound states of $N_S\sim 1/g^2$
particles~\cite{Drukier}. Then the suppression (\ref{eq:1}) is due to
multiparticle production~\cite{unitarity}.

In this Letter we propose general semiclassical method for computing
the leading suppression exponent $F(E)$ of the inclusive process ``two
particles $\to$ SA pair + particles.''  As a by--product, we
  calculate the exponent $F_N(E)$ of the same process with $N$ initial
  particles. In our method the problem is deformed by introducing a small
  parameter $\delta\rho$ which turns the process of SA pair production
  into a well--known tunneling process. To the best of our
  knowledge, no method of this kind has ever been proposed before.

\begin{figure}[t]
\includegraphics[width=0.99\columnwidth]{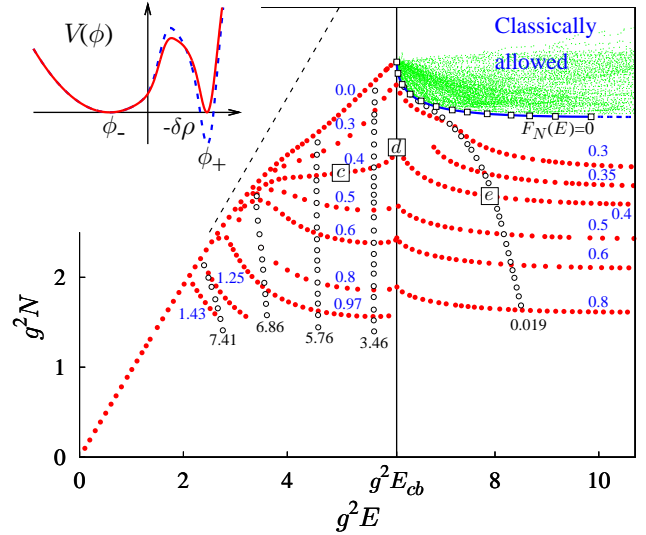}
\caption{\label{EN} Solutions in $(E,N)$ plane at
  $\delta\rho=0.4$. Numbers near the lines with empty and filled points
are the values of $T$ and $\theta$, respectively.
{\it Inset:} Potential density $V(\phi)$.}
\end{figure}

\begin{figure}[t]
\unitlength=0.01\columnwidth
\begin{picture}(100,90)
\put(-2,73){\includegraphics[width=.48\columnwidth]{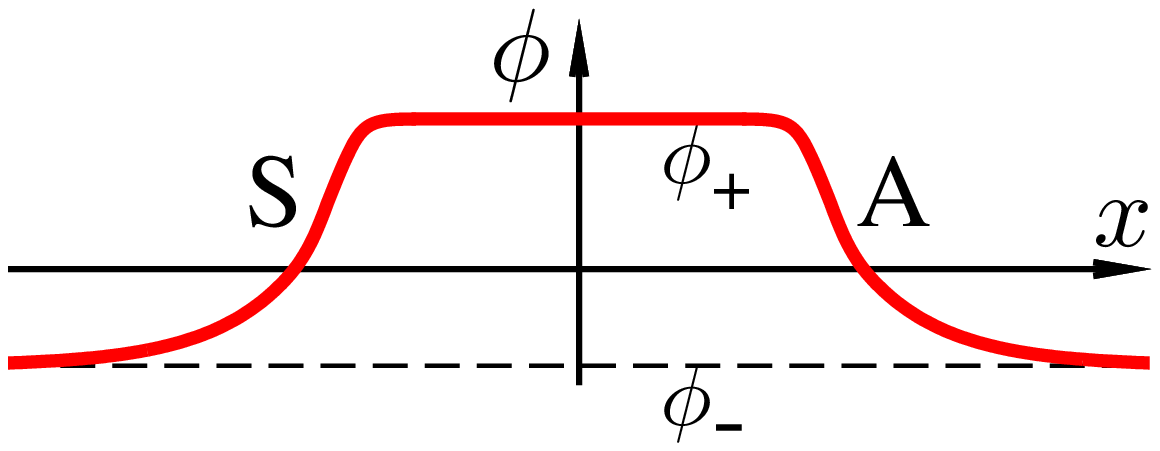}}
\put(3.3,85){\circle{3.5}}
\put(2.5,84.4){\small a}
\put(-2,52){\includegraphics[width=.48\columnwidth]{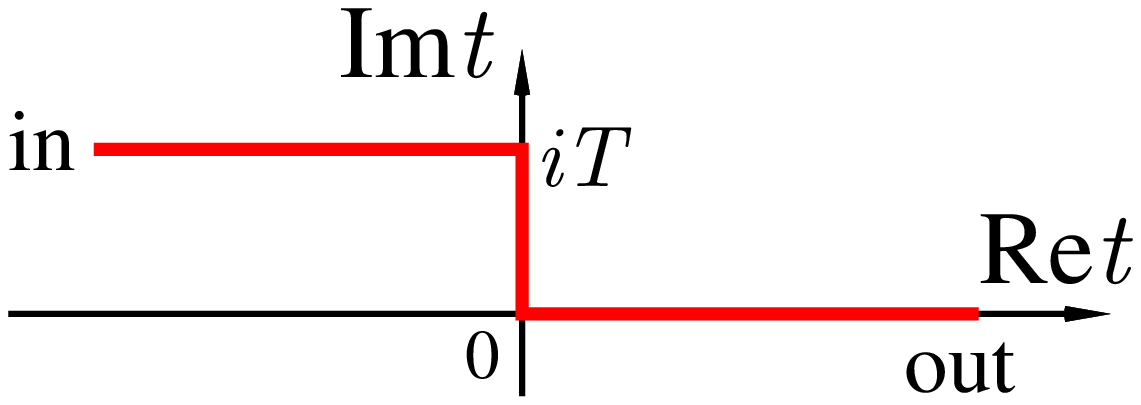}}
\put(3.3,68){\circle{3.5}}
\put(2.23,66.7){\small b}

\put(52,52){\includegraphics[width=.48\columnwidth]{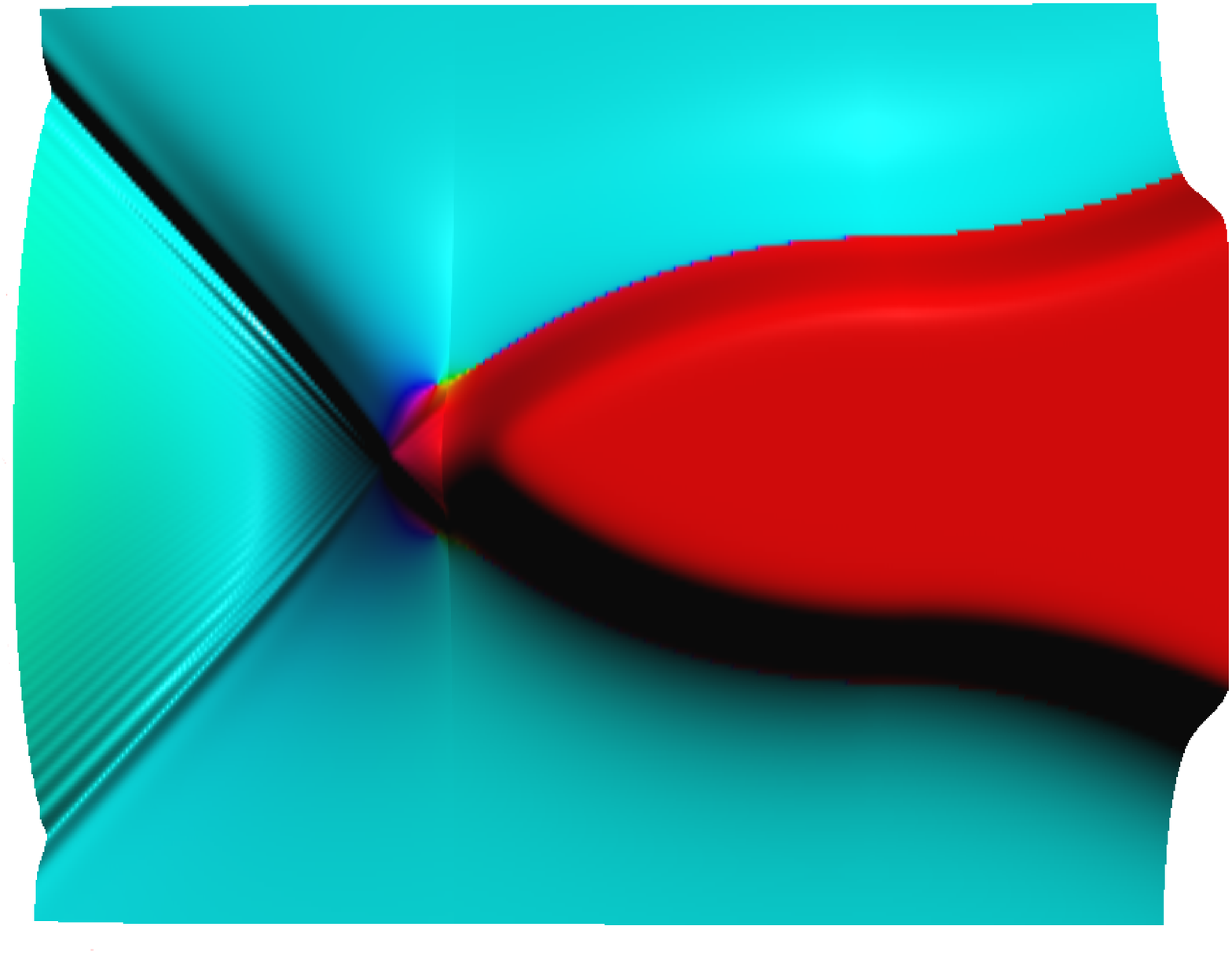}}
\put(51,70.5){\tiny 0}
\put(52,78){\tiny 2}
\put(52,84.8){\tiny 4}
\put(50.7,64){\tiny -2}
\put(51.3,56.5){\tiny -4}
\put(60.1,52){{\tiny -8}}
\put(66.4,52){{\tiny -6}}
\put(73.8,52){{\tiny -4}}
\put(79.3,52){{\tiny -2}}
\put(86,52){{\tiny 0}}
\put(92,52){{\tiny 2}}
\put(48.5,71){\begin{sideways}$x$\end{sideways}}
\put(68,48){Re$\,t-$Im$\,t$}
\put(67,85){\textcolor{white}{\circle*{3.5}}}
\put(66,84.5){\small c}

\put(-2,5){\includegraphics[width=.5\columnwidth]{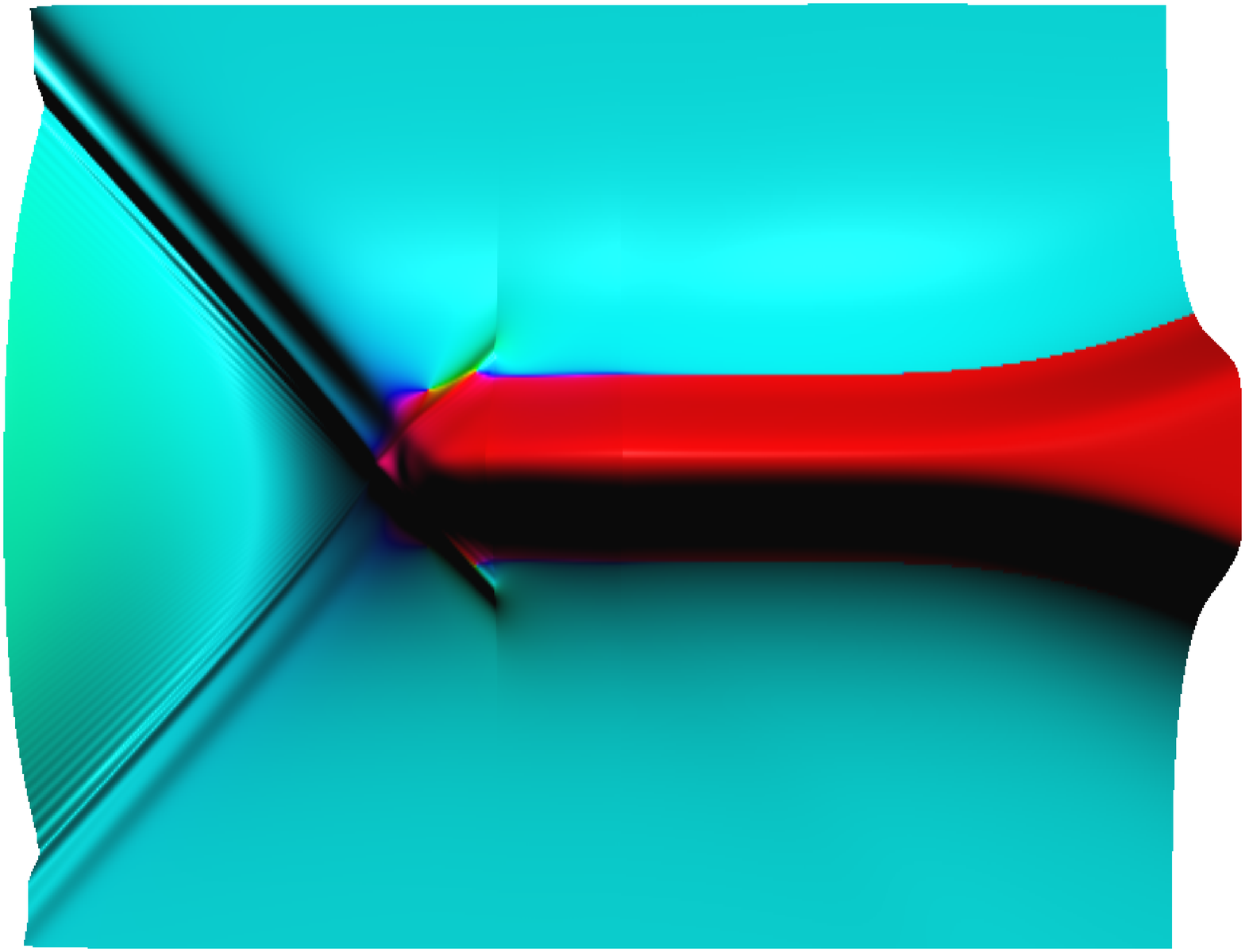}}
\put(-1.9,24.5){\tiny 0}
\put(-2,31.5){\tiny 2}
\put(-1.8,39){\tiny 4}
\put(-2.7,17){\tiny -2}
\put(-1.9,9.5){\tiny -4}
\put(3.7,5){{\tiny -6}}
\put(9.5,5){{\tiny -4}}
\put(16.5,5){{\tiny -2}}
\put(23.5,5){{\tiny 0}}
\put(30,5){{\tiny 2}}
\put(36.5,5){{\tiny 4}}
\put(43,5){{\tiny 6}}
\put(-4.7,23){\begin{sideways}$x$\end{sideways}}
\put(16,1){Re$\,t-$Im$\,t$}
\put(15,40){\textcolor{white}{\circle*{3.5}}}
\put(14,39.2){\small d}

\put(52,5){\includegraphics[width=.5\columnwidth]{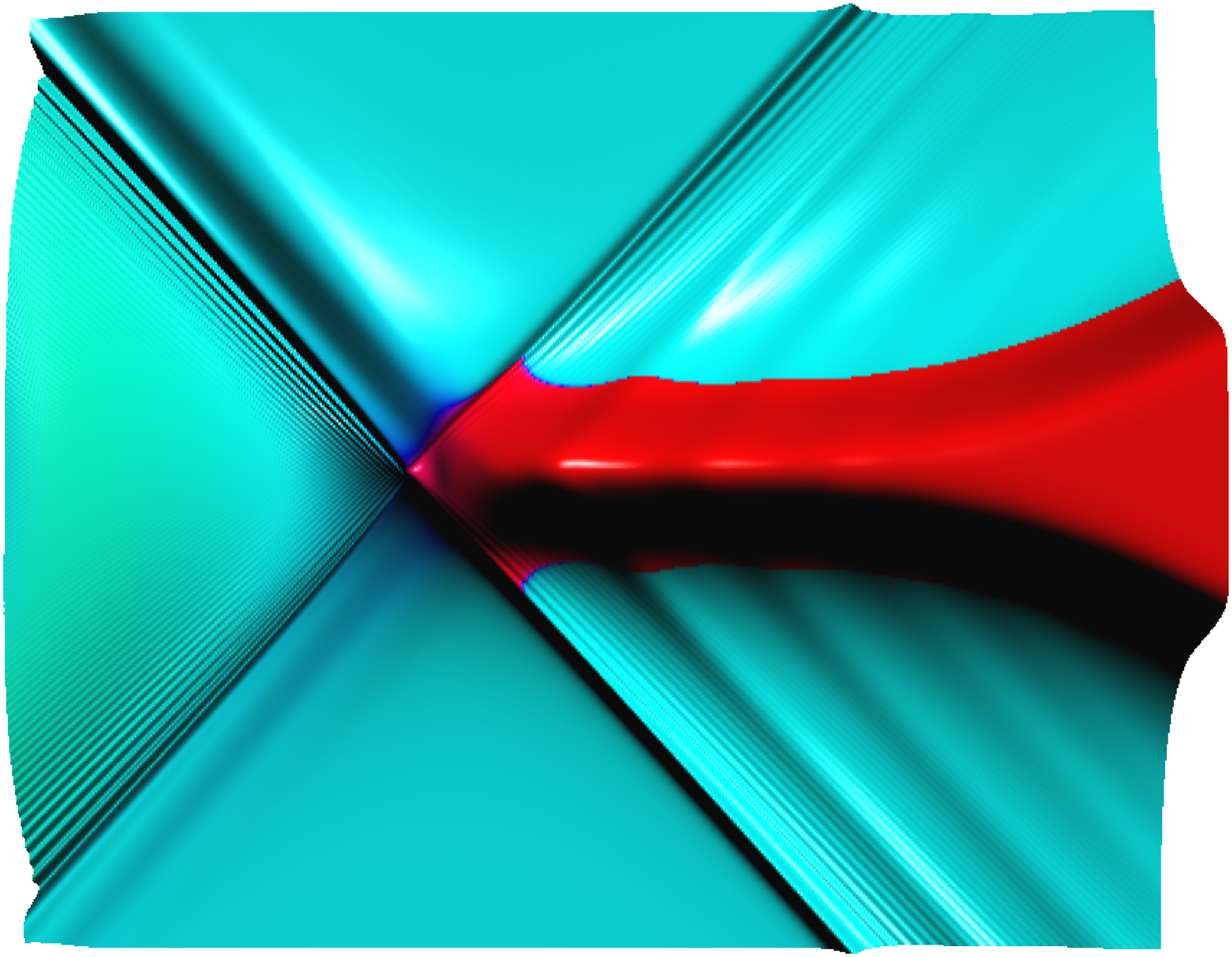}}
\put(51.8,24.7){\tiny 0}
\put(52,31.5){\tiny 2}
\put(52.8,39){\tiny 4}
\put(50.7,17){\tiny -2}
\put(50.9,9.5){\tiny -4}
\put(57.8,5){{\tiny -6}}
\put(63.5,5){{\tiny -4}}
\put(70.5,5){{\tiny -2}}
\put(77,5){{\tiny 0}}
\put(84,5){{\tiny 2}}
\put(90.5,5){{\tiny 4}}
\put(97,5){{\tiny 6}}
\put(49.5,23){\begin{sideways}$x$\end{sideways}}
\put(70,1){Re$\,t-$Im$\,t$}
\put(69,40){\textcolor{white}{\circle*{3.5}}}
\put(68,39.2){\small e}

\put(44,54.5){\includegraphics[width=1cm,angle=270]{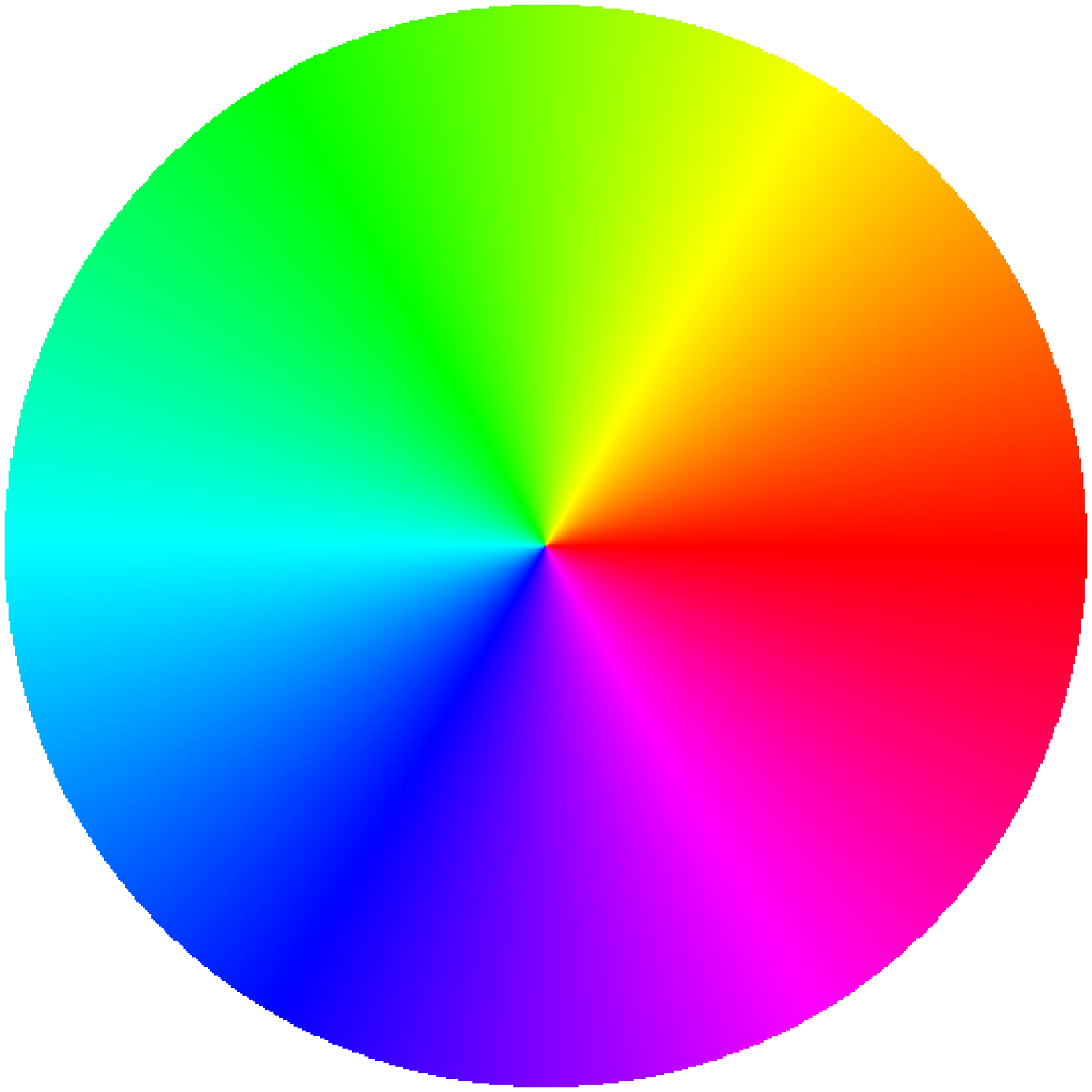}}
\put(55.7,47.5){\small 0}
\put(41.5,47.5){\small $\pi$}

\end{picture}
\caption{\label{solutions}(a) Soliton--antisoliton
  configuration. (b) Contour in complex time. (c)---(e) Semiclassical
  solutions $\phi_{s}(t,x)$ at $\delta\rho=0.4$; respective
  values of $(E,N)$ are shown by squares with
  letters in Fig.~\ref{EN}. In particular, (c) $E<E_{cb}$, (d)
  $E\approx E_{cb}$,  (e) 
  $E>E_{cb}$. Color represents $\mathrm{arg}\,\phi_s$.}
\end{figure}

For definiteness we consider $(1+1)$--dimensional scalar field theory
with action~\footnote{Upon field rescaling $\phi\to g\phi$,
 the coupling
  constant $g$ enters the non--linear terms in the potential only, 
as it should.}
\begin{equation}
\label{eq:2}S[\phi] = \frac{1}{g^2} \int d^2x \left[
-\frac12 \phi\, \Box \phi -
  V(\phi)\right]\;.
\end{equation}
This model possesses topological solitons if the scalar
potential  $V(\phi)$ has a pair of degenerate minima $\phi_-$
and $\phi_+$, see the inset in Fig.~\ref{EN}, solid
line. Soliton and antisoliton solutions interpolate between the
minima; their profiles are shown in Fig.~\ref{solutions}a.

An obstacle to the semiclassical description of SA pair
production is related to the fact that soliton and antisoliton
attract each other and annihilate classically into $N_{SA}\sim 1/g^2$
particles. Thus, there is no potential barrier separating SA pair from 
the particle sector and the process under study cannot be treated as
potential tunneling.

We get around this obstacle by introducing the potential barrier 
between SA pair and perturbative states. Namely, we add negative
energy density $(-\delta\rho)$ to the vacuum $\phi_+$, see  dashed
line in the inset in Fig.~\ref{EN}. This turns $\phi_-$ and $\phi_+$
into false and true vacua, respectively; the process of SA pair production
is now interpreted as false vacuum decay~\cite{false,*false1,*false2}
induced by particle collisions. The latter is a well--studied tunneling 
process~\cite{induced_thin_wall,*induced_thin_wall1,*induced_thin_wall2,%
*induced_thin_wall3,Kuznetsov}.
In the end of calculation we will take the limit $\delta \rho \to 0$.

The height of the potential barrier between the false and true vacua
is given by the energy $E_{cb}$ of the critical bubble~\cite{false}
--- unstable static solution ``sitting'' on top of the barrier. The
pressure $\delta \rho$ inside this bubble is balanced by the
soliton--antisoliton attraction. Let us estimate the critical bubble
size $d_{cb}$ at small $\delta \rho$. The attractive force $F_{att}$
between the soliton and antisoliton is proportional to the Yukawa
exponent $\exp(-m_+ d_{cb})$, where 
$m_+^2 = V''(\phi_+)$. Setting $F_{att} = \delta \rho$, one finds
${d_{cb} \sim -\log(\mbox{const} \cdot   \delta\rho) /m_+}$. We see
that at $\delta \rho \to 0$ the critical  bubble turns into a widely
separated SA pair and $E_{cb} \to 2M_S$, where $M_S$ is the soliton
mass.

Another difficulty is met in the case of $N=2$ particles in the
initial state, since states with few quanta cannot be described
semiclassically. We solve this problem by Rubakov--Son--Tinyakov (RST)
method~\cite{RST,*RST2},~\footnote{For alternative approach see
  S.~V.~Iordanskii and L.~P.~Pitaevskii, Sov. Phys. JETP {\bf 49}, 386
  (1979); S.~Yu.~Khlebnikov, Phys. Lett. B {\bf 282}, 459  (1992);
  D.~Diakonov, V.~Petrov, Phys. Rev. D {\bf 50}, 266 (1994).}. The
method is based on the assumption that as long as the number of colliding
quanta is semiclassically small, $N\ll 1/g^2$, the leading suppression
exponent $F_N(E)$ is universal, i.e. does not depend on $N$. In
detail, consider the $N$--particle inclusive probability,
\begin{equation}
\label{eq:3}
{\cal P}_N(E) = \sum_{i,f} \left|\langle i; E,N|\hat{\cal S}|f;
SA\rangle\right|^2 
\sim \mathrm{e}^{-F_N(E)/g^2}\;,
\end{equation}
where $\hat{\cal S}$ is S--matrix. The sums in Eq.~(\ref{eq:3}) run
over all initial states 
of energy $E$ and multiplicity $N$ and final states containing SA
pair and arbitrary number of quanta. 
RST conjecture states that the multiparticle suppression
exponent $F_N(E)$ in Eq.~(\ref{eq:3}) coincides with the two--particle
exponent $F(E)$ for $N\ll 1/g^2$. On the other hand, at $N \gg 1$ the
initial states in Eq.~(\ref{eq:3}) can be described
semiclassically. Thus, $F(E)$ is computed by evaluating semiclassically
$F_N(E)$ and taking the small--$N$ limit
\begin{equation}
\label{eq:4}
F(E) = \lim_{g^2 N\to 0} F_N(E)\;.
\end{equation}
It is worth noting that the conjecture (\ref{eq:4}) has been supported by
field theory 
calculations~\cite{RSTcheck,*RSTcheck1} and proved in the context of
multidimensional quantum mechanics~\cite{Bonini,LPS}. 

After modifying the original problem we arrived at false vacuum
decay induced by collision of $N\gg 1$ particles. The latter process
can be described semiclassically, as outlined below. The price to pay
is the limits $\delta\rho \to 0$ and $g^2N \to 0$, which
should be taken in the end of calculation. In particular, we are going
to show that the limit $\delta \rho \to 0$ of the suppression exponent
exists.

Semiclassical method for the calculation of the multiparticle probability
(\ref{eq:3}) at $N\gg 1$ and $g \ll 1$ has been developed in 
Ref.~\cite{RST}. This  
method is based on the saddle-point evaluation~\footnote{The
  saddle--point method is justified at small $g^2$ when the integrand
  $\exp(iS[\phi])$ in the path integral rapidly oscillates,
  cf. Eq.~(\ref{eq:2}).} of the path integral for the
probability. Below we list the boundary conditions for the complex
saddle-point configuration $\phi_{s}(t,x) \in \mathbb{C}$ and give
expression for $F_N(E)$; see 
Refs.~[\onlinecite{RST}, \onlinecite{Kuznetsov}] for derivation.

Configuration $\phi_{s}(t,x)$ satisfies classical field equation
$\delta S/\delta \phi = 0$ along the contour in complex time shown in 
Fig.~\ref{solutions}b, where the Euclidean part corresponds to
tunneling. In the asymptotic past $\phi_{s}(t,x)$ is
a collection of linear waves above the false vacuum $\phi_-$,
\begin{align}
\notag
\phi_{s} \to \phi_- + \int
&\frac{dk}{\sqrt{2\omega_k}} \left[a_k \mathrm{e}^{-i\omega_k t +ikx} +
  b_k^* \mathrm{e}^{i\omega_k t -ikx}\right]\\
\label{eq:12}
&\qquad\qquad\qquad\mbox{as} \;\; t\to iT-\infty\;,
\end{align}
where $\omega_k^2 = k^2 + m_-^2$, $m_-^2 =
V''(\phi_-)$. The first boundary condition is relation between the
amplitudes~\cite{RST},
\begin{subequations}
\label{eq:11}
\begin{equation}
\label{eq:6}
a_k = \mathrm{e}^{-2\omega_k T-\theta} b_k\;,
\end{equation}
where $T$ and $\theta$ are real parameters.
In the asymptotic future $\phi_{s}(t,x)$ contains a bubble of true
vacuum. The second boundary condition is asymptotic real--valuedness:
\begin{equation}
\label{eq:7}
\mathrm{Im}\, \phi_{s}, \; \mathrm{Im}\, \partial_t\phi_{s} \to 0 \qquad
\mbox{as}\;\; t\to +\infty\;.
\end{equation}
\end{subequations}
Equations (\ref{eq:11}) are sufficient to specify complex
solution $\phi_{s}(t,x)$ for given values of $T$ and $\theta$ in
Eq.~(\ref{eq:6}).

Parameters $T$ and $\theta$ are related to $(E,N)$
by the saddle--point conditions~[\onlinecite{RST}, \onlinecite{Kuznetsov}]
\begin{equation}
\label{eq:9}
g^2 E = {2\pi} \int dk \,\omega_k  a_k b_k^*\;,\qquad 
g^2 N = {2\pi} \int dk \,a_k b_k^*\;.
\end{equation}
Given the saddle--point configuration $\phi_s(t,x)$,
one evaluates the suppression exponent,
\begin{equation}
\label{eq:8}
F_N(E) = g^2\left(2\mathrm{Im}\, S[\phi_{s}] - 2ET -
N\theta\right)\;,
\end{equation}
where the last two terms are due to non-trivial initial state, see
Ref.~\cite{Kuznetsov}. Note that $\phi_s(t,x)$ and $F_N(E)$ depend on
$g$, $E$, $N$ only via combinations $g^2 N$, $g^2 E$, see
Eqs.~(\ref{eq:9}).

To summarize, the recipe for calculating the suppression
exponent of SA pair production in two--particle collisions is as follows. 
One starts at $\delta \rho>0$ by finding two--parametric family of
saddle--point configurations $\phi_{s}(t,x)$ which satisfy the
classical field  equation and boundary conditions (\ref{eq:11}).
Then one computes the values of $E$, $N$ and
$F_N(E)$, Eqs.~(\ref{eq:9}), (\ref{eq:8}), for each of these
configurations. The result for the suppression exponent $F(E)$ in
Eq.~(\ref{eq:1}) is recovered in the limits 
$\delta\rho\to 0$ and $g^2 N \to 0$. 

We support the method by performing explicit calculations in the model
(\ref{eq:2}) with potential shown in the inset in Fig.~\ref{EN},
\begin{equation}
\label{eq:10}
V(\phi) = \frac{1}{2}(\phi+1)^2\left[1 - v\,
  W\left(\frac{\phi-1}{a}\right)\right]\;,
\end{equation}
where $W(x) = \mathrm{e}^{-x^2}( x + x^3 + x^5)$, $a=0.4$ and $v$ is
tuned to provide the required value of $\delta \rho$. We do not use the
standard $\phi^4$ potential because of the chaotic properties
of $\phi^4$ kink--antikink
dynamics~[\onlinecite{kink_chaos}, \onlinecite{solitons}] which lead to
difficulties in the semiclassical analysis~\cite{chaos_semiclassical,%
*chaos_semiclassical1,*chaos_semiclassical2,Onishi,*Onishi1}.

We solve the semiclassical boundary value problem
    numerically~\footnote{We discretize Eqs.~(\ref{eq:11}), 
    (\ref{eq:9}), (\ref{eq:8}) by introducing uniform lattice
    $N_t\times N_x = 5000\times 250$ of extent $L_t\times L_x =
    15\times 7$. Precision of discretization is kept smaller than
    $0.1\%$; it is controlled by changing
    the lattice size and keeping track of energy conservation.
    Linear evolution in Eq.~(\ref{eq:12})
    holds with accuracy smaller than $1\%$. Extrapolation
    $\delta\rho \to 0$ produces relative errors in $F(E)$ of order
    $10^{-3}$.}  using 
methods of
Refs.~\cite{Kuznetsov,SU2,*SU21}. Our starting point 
is bounce~\cite{false} --- Euclidean solution describing spontaneous
decay of false vacuum at $E=N=0$. By using the classical field
equation we continue the bounce to the Minkowski parts of the contour
in 
Fig.~\ref{solutions}b. Then, changing $(T,\theta)$ (and thus $(E,N)$)
in small steps and solving  numerically the boundary value problem 
(\ref{eq:11}),
we construct the continuous family of saddle--point configurations
$\phi_{s}(t,x)$ at $E<E_{cb}$. Each configuration is represented by
a point in the left part of Fig.~\ref{EN}. The points form the
lines $\theta=\mbox{const}$ (filled points) and $T=\mbox{const}$
(empty points). 

Solutions at $E< E_{cb}$, $E_{cb}\approx 2M_S$ have the form of
distorted bounces, see Fig.~\ref{solutions}c. Wave packets in the left
part of the figure represent particles moving in the initial state;
after collision, the particles back--react on the Euclidean part of
solution. Using the semiclassical solutions, we calculate the
multiparticle exponent $F_N(E)$ at $E<E_{cb}$.

We evaluate the two--particle exponent $F(E)$ by extrapolating
$F_N(E)$~\footnote{One can argue~\cite{RST,Levkov} that $g^2N
  \cdot\theta\to 0$ as $g^2 N\to 0$, so we use only the first two
  terms in Eq.~(\ref{eq:8}).} to $g^2 N = 0$ with quadratic
polynomials in $g^2 N$, cf. Eq.~(\ref{eq:4}). The accuracy of
extrapolation is~5\%.
 
The probability (\ref{eq:1}) of SA pair production must vanish below
the kinematic threshold $E=2M_S$. Let us show that, indeed,  $F(E)\to
+\infty$  as $\delta \rho\to 0$ in the region $E<E_{cb}\to 2M_S$. We
recall that at small $\delta \rho$ the thin--wall
approximation~\cite{false} is applicable and the two--particle
exponent $F(E)$ can be evaluated analytically modulo
$O(\delta\rho^0)$ corrections~\cite{induced_thin_wall},
\begin{equation} 
\label{eq:5}
F = \frac{g^4M_{S}^{2}}{\delta\rho}\left(2\arccos{\frac{E}{2M_{S}}} -
\frac{E}{M_{S}}\sqrt{1-\frac{E^{2}}{4M_{S}^{2}}}\right)\;.
\end{equation}
We see that $F(E)\propto 1/\delta\rho$ at $E<2M_S$; this property
disappears for $E\geq 2M_S$ where the thin--wall approximation breaks
down.

\begin{figure}[t]
\centerline{\includegraphics[width=\columnwidth]{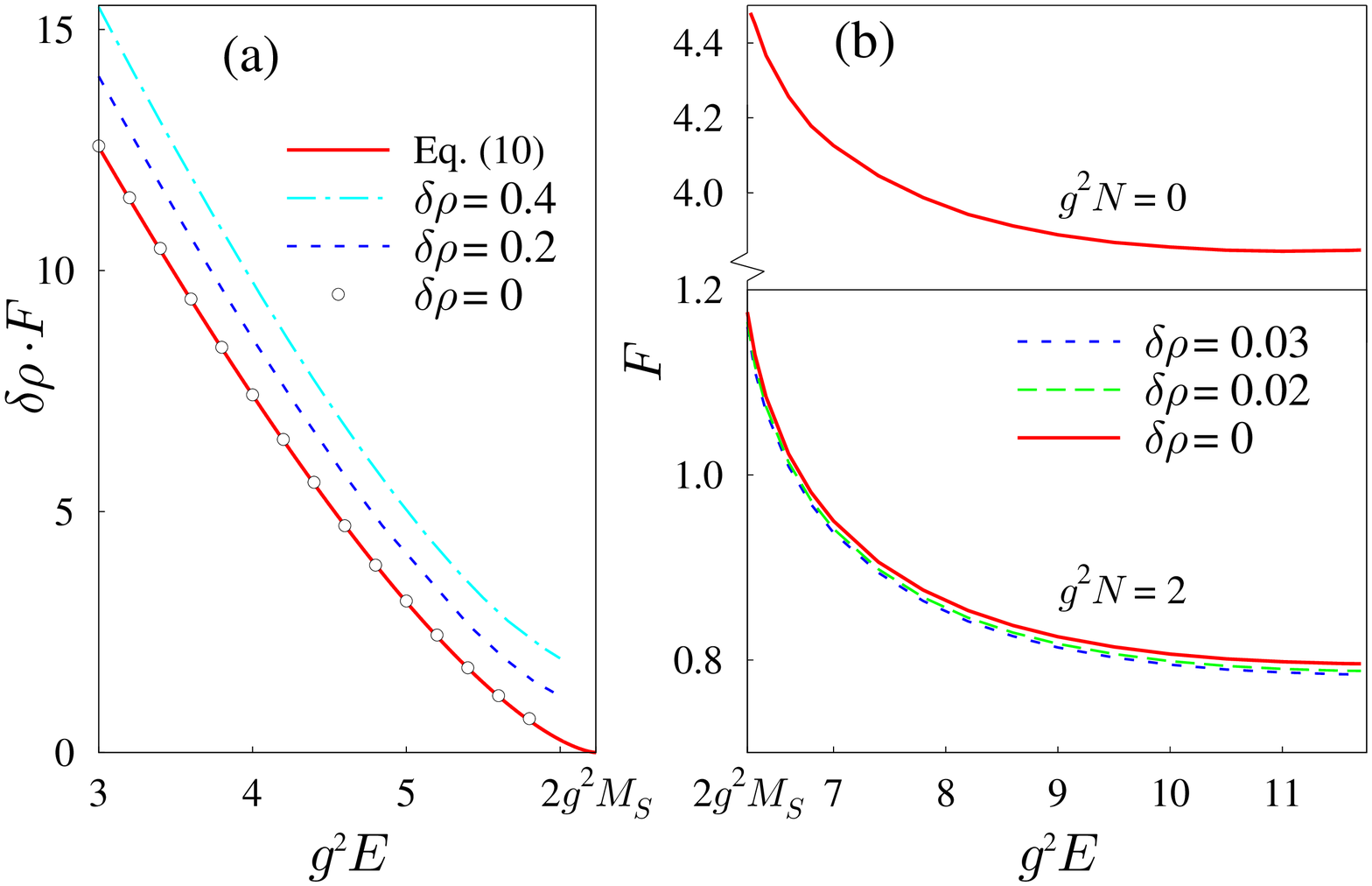}}
\caption{\label{F} (a) Function $\delta \rho \cdot F(E)$ at $g^2N=0$, $E<2M_S$.
(b) Suppression exponent $F_N(E)$ at $E>2M_S$.} 
\end{figure}

Our numerical results for $\delta\rho \,\cdot F(E)$ at $g^2 N=0$
(Fig.~\ref{F}a, dashed lines) approach Eq.~(\ref{eq:5})
(Fig.~\ref{F}a, solid line) as
$\delta \rho$ decreases and coincide with it after extrapolation to
$\delta\rho= 0$ (points). This gives support to our method.

The fact that $\delta \rho\, \cdot F(E) \to 0$ as $E\to 2M_S$, $\delta
\rho \to 0$
hints that the properties of semiclassical solutions become
qualitatively different at $E>E_{cb}$. This is indeed the case;
in
fact, our numerical procedure does not produce solutions 
at $E>E_{cb}$ at all. By inspecting $\phi_s(t,x)$ with $E\approx E_{cb}$,
Fig.~\ref{solutions}d, we see the reason for that: this solution
has long, almost static part where it is close to the
critical bubble. The instability of the latter makes the numerical
techniques inefficient.

We solve the instability problem by $\epsilon$--regularization
method of Refs.~[\onlinecite{epsilon,*epsilon1,*epsilon2},
  \onlinecite{LPS}] which prescribes to add small imaginary term to
the potential, 
$V(\phi) \to V_{\epsilon}(\phi,x) = V(\phi) - i \epsilon f(\phi,x)$, 
where $f(\phi,x) \geq 0$ at real $\phi$, ${f(\phi_+,x)=0}$. For
$\epsilon>0$ the critical bubble has complex energy and cannot be
approached by $\phi_{s}(t,x)$: the energy of the latter is real due to
Eq.~(\ref{eq:7}). Thus, the
semiclassical solutions are no longer unstable
at $\epsilon>0$ and we are able to cover the region
$E>E_{cb}$ with new solutions, see Fig.~\ref{EN}. Using
Eq.~(\ref{eq:8}) and taking the limit $\epsilon \to 0$, we compute the
multiparticle suppression exponent $F_N(E)$.
The typical saddle--point configuration at $E>E_{cb}$ is shown in
Fig.~\ref{solutions}e. It still describes formation of the critical
bubble~\footnote{The critical bubble decays in the right part of
  Fig.~\ref{solutions}e due to regularization $\epsilon>0$.}; the
energy excess $E-E_{cb}$ is radiated away in the middle 
part of the solution. 

It is worth noting  that formation of classically unstable ``states'' is a
manifestation of general tunneling mechanism proposed recently in
multidimensional quantum mechanics~[\onlinecite{Onishi},
  \onlinecite{epsilon},
  \onlinecite{new_mechanism,*new_mechanism1},
  \onlinecite{LPS}] and quantum field
theory~\cite{SU2,Levkov,*Levkov1}.

To check the semiclassical method, we find values of $(E,N)$
  corresponding to classically allowed decay of false
  vacuum~\cite{Demidov}. On the one hand, these values are obtained 
  from the semiclassical exponent, since classical transitions
correspond to $F_N(E)=0$. Figure~\ref{EN} shows the line $F_N(E)=0$
obtained in this way. On the other hand, ``classically allowed''
values of $(E,N)$ can be obtained by studying classical  
solutions~[\onlinecite{Rebbi}]. To
this end, we performed~\cite{Demidov} Monte Carlo simulation over
the sets of Cauchy data $\{\phi(0,x),\, \dot{\phi}(0,x)\}$. Starting
from each set, we solved numerically the classical field equation and
obtained solution $\phi_{cl}(t,x)$. We  selected solutions describing
transitions between the false and true vacua and computed the values
  of $(E,N)$, Eqs.~(\ref{eq:12}), (\ref{eq:9}), for these solutions.
The latter values are shown in Fig.~\ref{EN} by
dots which fill precisely the region bounded by our line
$F_N(E)=0$. We conclude that the semiclassical result for $F_N(E)$ is
reliable.
 
The exponent $F_N(E)$ is plotted in Fig.~\ref{F}b at $N=2/g^2$ and
different values of $\delta\rho$ (dashed lines). The graphs are
almost indistinguishable; thus, the limit ${\delta\rho\to 0}$
exists. Extrapolating $F_N(E)$ to $\delta\rho=0$, we obtain the final
result for the suppression exponent of SA pair production in
$N$--particle collisions (solid line in Fig.~\ref{F}b). The
two--particle exponent $F(E)$ is recovered by extrapolating $F_N(E)$ to
$g^2 N=0$ (upper graph in Fig.~\ref{F}b). 

At high energies our suppression exponents are almost constant which
is a typical behavior for collision--induced
suppressions~[\onlinecite{Shifman}, \onlinecite{SU2},
  \onlinecite{Levkov}]. We expect that this feature holds in other
models.

We end up by arguing that the method proposed in this Letter is 
applicable in 
multidimensional ($D>2$) and gauge theories. The problem with
soliton--antisoliton attraction can be solved in the general
case by introducing  
small constant force ${\bf F}$, an analog of $\delta \rho$,  which
drags soliton and antisoliton apart. At $|{\bf F}|\ne 0$ SA pair is
separated from the perturbative sector by a potential barrier;
tunneling through this barrier is a Schwinger process of
spontaneous soliton--antisoliton pair creation in the external
field~\cite{Manton,*Monopoles}.  For
example, in the case of  t'Hooft--Polyakov monopoles the external
force is introduced by adding uniform magnetic field. Other steps of
the method --- RST conjecture (\ref{eq:4}) and semiclassical equations
(\ref{eq:11}),~(\ref{eq:9}), (\ref{eq:8}) --- are explicitly general,
cf. Ref.~[\onlinecite{RST}]. In particular, suppression exponent of
Schwinger process is proportional to $1/|{\bf
  F}|$~\cite{Manton,*Monopoles}; this enhancement
vanishes~\cite{schwinger_induced} at $E=2M_S$. The final result for
the semiclassical exponent of collision--induced SA pair production
should be obtained by taking the limit $|{\bf F}|\to 0$ at $E>2M_S$.

We thank V.A~Rubakov, S.M.~Sibiryakov, F.L.~Bezrukov and G.I.~Rubtsov
for discussions. This work was supported in part by grants
NS-5525.2010.2; MK-7748.2010.2, grant of the ``Dynasty'' foundation
(D.L.); RFBR-11-02-01528-a and Russian state contracts
02.740.11.0244, P520, P2598 (S.D.). Numerical calculations have
been performed on the Computational cluster of the Theoretical
division of INR RAS.
\bibliography{articles}

%Merlin.mbs v4.21 2009-07-09.
\begin{thebibliography}{10}%
\makeatletter
\providecommand \@ifxundefined [1]{%
 \ifx #1\undefined \expandafter \@firstoftwo
 \else \expandafter \@secondoftwo
\fi
}%
\providecommand \@ifnum [1]{%
 \ifnum #1\expandafter \@firstoftwo
 \else \expandafter \@secondoftwo
\fi
}%
\providecommand \enquote [1]{``#1''}%
\providecommand \bibnamefont  [1]{#1}%
\providecommand \bibfnamefont [1]{#1}%
\providecommand \citenamefont [1]{#1}%
\providecommand\href[0]{\@sanitize\@href}%
\providecommand\@href[1]{\endgroup\@@startlink{#1}\endgroup\@@href}%
\providecommand\@@href[1]{#1\@@endlink}%
\providecommand \@sanitize [0]{\begingroup\catcode`\&12\catcode`\#12\relax}%
\@ifxundefined \pdfoutput {\@firstoftwo}{%
 \@ifnum{\z@=\pdfoutput}{\@firstoftwo}{\@secondoftwo}%
}{%
 \providecommand\@@startlink[1]{\leavevmode\special{html:<a href="#1">}}%
 \providecommand\@@endlink[0]{\special{html:</a>}}%
}{%
 \providecommand\@@startlink[1]{%
  \leavevmode
  \pdfstartlink
   attr{/Border[0 0 1 ]/H/I/C[0 1 1]}%
   user{/Subtype/Link/A<</Type/Action/S/URI/URI(#1)>>}%
  \relax
 }%
 \providecommand\@@endlink[0]{\pdfendlink}%
}%
\providecommand \url  [0]{\begingroup\@sanitize \@url }%
\providecommand \@url [1]{\endgroup\@href {#1}{\urlprefix}}%
\providecommand \urlprefix [0]{URL }%
\providecommand \Eprint[0]{\href }%
\@ifxundefined \urlstyle {%
  \providecommand \doi [1]{doi:\discretionary{}{}{}#1}%
}{%
  \providecommand \doi [0]{doi:\discretionary{}{}{}\begingroup
  \urlstyle{rm}\Url }%
}%
\providecommand \doibase [0]{http://dx.doi.org/}%
\providecommand \Doi[1]{\href{\doibase#1}}%
\providecommand \bibAnnote [3]{%
  \BibitemShut{#1}%
  \begin{quotation}\noindent
    \textsc{Key:}\ #2\\\textsc{Annotation:}\ #3%
  \end{quotation}%
}%
\providecommand \bibAnnoteFile [2]{%
  \IfFileExists{#2}{\bibAnnote {#1} {#2} {\input{#2}}}{}%
}%
\providecommand \typeout [0]{\immediate \write \m@ne }%
\providecommand \selectlanguage [0]{\@gobble}%
\providecommand \bibinfo [0]{\@secondoftwo}%
\providecommand \bibfield [0]{\@secondoftwo}%
\providecommand \translation [1]{[#1]}%
\providecommand \BibitemOpen[0]{}%
\providecommand \bibitemStop [0]{}%
\providecommand \bibitemNoStop [0]{.\EOS\space}%
\providecommand \EOS [0]{\spacefactor3000\relax}%
\providecommand \BibitemShut [1]{\csname bibitem#1\endcsname}%
%</preamble>
\bibitem{Drukier}%
  \BibitemOpen
  \bibfield{author}{%
  \bibinfo {author} {\bibfnamefont{A.~K.}\ \bibnamefont{Drukier}}\ and\
  \bibinfo {author} {\bibfnamefont{S.}~\bibnamefont{Nussinov}},\ }%
  \bibfield{journal}{%
  \bibinfo {journal} {Phys. Rev. Lett.}\ }%
  \textbf{\bibinfo {volume} {49}},\ \bibinfo {pages} {102} (\bibinfo {year}
  {1982})%
  \bibAnnoteFile{NoStop}{Drukier}%
\bibitem{solitons}%
  \BibitemOpen
  \bibfield{author}{%
  \bibinfo {author} {\bibfnamefont{S.}~\bibnamefont{Dutta}}, \bibinfo {author}
  {\bibfnamefont{D.~A.}\ \bibnamefont{Steer}},\ and\ \bibinfo {author}
  {\bibfnamefont{T.}~\bibnamefont{Vachaspati}},\ }%
  \bibfield{journal}{%
  \bibinfo {journal} {Phys. Rev. Lett.}\ }%
  \textbf{\bibinfo {volume} {101}},\ \bibinfo {pages} {121601} (\bibinfo {year}
  {2008})%
  \bibAnnoteFile{NoStop}{solitons}%
\bibitem{solitons1}%
  \BibitemOpen
  \bibfield{author}{%
  \bibinfo {author} {\bibfnamefont{T.}~\bibnamefont{Romanczukiewicz}}\ and\
  \bibinfo {author} {\bibfnamefont{Y.}~\bibnamefont{Shnir}},\ }%
  \bibfield{journal}{%
  \bibinfo {journal} {Phys. Rev. Lett.}\ }%
  \textbf{\bibinfo {volume} {105}},\ \bibinfo {pages} {081601} (\bibinfo {year}
  {2010})%
  \bibAnnoteFile{NoStop}{solitons1}%
\bibitem{unitarity}%
  \BibitemOpen
  \bibfield{author}{%
  \bibinfo {author} {\bibfnamefont{T.}~\bibnamefont{Banks}}, \bibinfo {author}
  {\bibfnamefont{G.}~\bibnamefont{Farrar}}, \bibinfo {author}
  {\bibfnamefont{M.}~\bibnamefont{Dine}}, \bibinfo {author}
  {\bibfnamefont{D.}~\bibnamefont{Karabali}},\ and\ \bibinfo {author}
  {\bibfnamefont{B.}~\bibnamefont{Sakita}},\ }%
  \bibfield{journal}{%
  \bibinfo {journal} {Nucl. Phys. B}\ }%
  \textbf{\bibinfo {volume} {347}},\ \bibinfo {pages} {581 } (\bibinfo {year}
  {1990})%
  \bibAnnoteFile{NoStop}{unitarity}%
\bibitem{unitarity1}%
  \BibitemOpen
  \bibfield{author}{%
  \bibinfo {author} {\bibfnamefont{V.~I.}\ \bibnamefont{Zakharov}},\ }%
  \bibfield{journal}{%
  \bibinfo {journal} {Phys. Rev. Lett.}\ }%
  \textbf{\bibinfo {volume} {67}},\ \bibinfo {pages} {3650} (\bibinfo {year}
  {1991})%
  \bibAnnoteFile{NoStop}{unitarity1}%
\bibitem{unitarity2}%
  \BibitemOpen
  \bibfield{author}{%
  \bibinfo {author} {\bibfnamefont{H.}~\bibnamefont{Goldberg}},\ }%
  \bibfield{journal}{%
  \bibinfo {journal} {Phys. Lett. B}\ }%
  \textbf{\bibinfo {volume} {257}},\ \bibinfo {pages} {346 } (\bibinfo {year}
  {1991})%
  \bibAnnoteFile{NoStop}{unitarity2}%
\bibitem{Note1}%
  \BibitemOpen
  \bibinfo {note} {Upon field rescaling $\phi \to g\phi $, the coupling
  constant $g$ enters the non--linear terms in the potential only, as it
  should.}%
  \bibAnnoteFile{Stop}{Note1}%
\bibitem{false}%
  \BibitemOpen
  \bibfield{author}{%
  \bibinfo {author} {\bibfnamefont{M.~B.}\ \bibnamefont{Voloshin}}, \bibinfo
  {author} {\bibfnamefont{I.~Y.}\ \bibnamefont{Kobzarev}},\ and\ \bibinfo
  {author} {\bibfnamefont{L.~B.}\ \bibnamefont{Okun}},\ }%
  \bibfield{journal}{%
  \bibinfo {journal} {Sov. J. Nucl. Phys.}\ }%
  \textbf{\bibinfo {volume} {20}},\ \bibinfo {pages} {644} (\bibinfo {year}
  {1975})%
  \bibAnnoteFile{NoStop}{false}%
\bibitem{false1}%
  \BibitemOpen
  \bibfield{author}{%
  \bibinfo {author} {\bibfnamefont{M.}~\bibnamefont{Stone}},\ }%
  \bibfield{journal}{%
  \bibinfo {journal} {Phys. Lett. B}\ }%
  \textbf{\bibinfo {volume} {67}},\ \bibinfo {pages} {186 } (\bibinfo {year}
  {1977})%
  \bibAnnoteFile{NoStop}{false1}%
\bibitem{false2}%
  \BibitemOpen
  \bibfield{author}{%
  \bibinfo {author} {\bibfnamefont{S.}~\bibnamefont{Coleman}},\ }%
  \bibfield{journal}{%
  \bibinfo {journal} {Phys. Rev. D}\ }%
  \textbf{\bibinfo {volume} {15}},\ \bibinfo {pages} {2929} (\bibinfo {year}
  {1977})%
  \bibAnnoteFile{NoStop}{false2}%
\bibitem{induced_thin_wall}%
  \BibitemOpen
  \bibfield{author}{%
  \bibinfo {author} {\bibfnamefont{M.}~\bibnamefont{Voloshin}}\ and\ \bibinfo
  {author} {\bibfnamefont{K.}~\bibnamefont{Selivanov}},\ }%
  \bibfield{journal}{%
  \bibinfo {journal} {JETP Lett.}\ }%
  \textbf{\bibinfo {volume} {42}},\ \bibinfo {pages} {422} (\bibinfo {year}
  {1985})%
  \bibAnnoteFile{NoStop}{induced_thin_wall}%
\bibitem{induced_thin_wall1}%
  \BibitemOpen
  \bibfield{author}{%
  \bibinfo {author} {\bibfnamefont{M.}~\bibnamefont{Voloshin}}\ and\ \bibinfo
  {author} {\bibfnamefont{K.}~\bibnamefont{Selivanov}},\ }%
  \bibfield{journal}{%
  \bibinfo {journal} {Sov. J. Nucl. Phys.}\ }%
  \textbf{\bibinfo {volume} {44}},\ \bibinfo {pages} {868} (\bibinfo {year}
  {1986})%
  \bibAnnoteFile{NoStop}{induced_thin_wall1}%
\bibitem{induced_thin_wall2}%
  \BibitemOpen
  \bibfield{author}{%
  \bibinfo {author} {\bibfnamefont{V.~A.}\ \bibnamefont{Rubakov}}, \bibinfo
  {author} {\bibfnamefont{D.~T.}\ \bibnamefont{Son}},\ and\ \bibinfo {author}
  {\bibfnamefont{P.~G.}\ \bibnamefont{Tinyakov}},\ }%
  \bibfield{journal}{%
  \bibinfo {journal} {Phys. Lett. B}\ }%
  \textbf{\bibinfo {volume} {278}},\ \bibinfo {pages} {279 } (\bibinfo {year}
  {1992})%
  \bibAnnoteFile{NoStop}{induced_thin_wall2}%
\bibitem{induced_thin_wall3}%
  \BibitemOpen
  \bibfield{author}{%
  \bibinfo {author} {\bibfnamefont{V.~G.}\ \bibnamefont{Kiselev}},\ }%
  \bibfield{journal}{%
  \bibinfo {journal} {Phys. Rev. D}\ }%
  \textbf{\bibinfo {volume} {45}},\ \bibinfo {pages} {2929} (\bibinfo {year}
  {1992})%
  \bibAnnoteFile{NoStop}{induced_thin_wall3}%
\bibitem{Kuznetsov}%
  \BibitemOpen
  \bibfield{author}{%
  \bibinfo {author} {\bibfnamefont{A.~N.}\ \bibnamefont{Kuznetsov}}\ and\
  \bibinfo {author} {\bibfnamefont{P.~G.}\ \bibnamefont{Tinyakov}},\ }%
  \bibfield{journal}{%
  \bibinfo {journal} {Phys. Rev. D}\ }%
  \textbf{\bibinfo {volume} {56}},\ \bibinfo {pages} {1156} (\bibinfo {year}
  {1997})%
  \bibAnnoteFile{NoStop}{Kuznetsov}%
\bibitem{RST}%
  \BibitemOpen
  \bibfield{author}{%
  \bibinfo {author} {\bibfnamefont{V.~A.}\ \bibnamefont{Rubakov}}, \bibinfo
  {author} {\bibfnamefont{D.~T.}\ \bibnamefont{Son}},\ and\ \bibinfo {author}
  {\bibfnamefont{P.~G.}\ \bibnamefont{Tinyakov}},\ }%
  \bibfield{journal}{%
  \bibinfo {journal} {Phys. Lett. B}\ }%
  \textbf{\bibinfo {volume} {287}},\ \bibinfo {pages} {342 } (\bibinfo {year}
  {1992})%
  \bibAnnoteFile{NoStop}{RST}%
\bibitem{RST2}%
  \BibitemOpen
  \bibfield{author}{%
  \bibinfo {author} {\bibfnamefont{V.~A.}\ \bibnamefont{Rubakov}}\ and\
  \bibinfo {author} {\bibfnamefont{M.~E.}\ \bibnamefont{Shaposhnikov}},\ }%
  \bibfield{journal}{%
  \bibinfo {journal} {Phys. Usp.}\ }%
  \textbf{\bibinfo {volume} {39}},\ \bibinfo {pages} {461} (\bibinfo {year}
  {1996})%
  \bibAnnoteFile{NoStop}{RST2}%
\bibitem{Note2}%
  \BibitemOpen
  \bibinfo {note} {For alternative approach see S.~V.~Iordanskii and
  L.~P.~Pitaevskii, Sov. Phys. JETP {\protect \bf 49}, 386 (1979);
  S.~Yu.~Khlebnikov, Phys. Lett. B {\protect \bf 282}, 459 (1992); D.~Diakonov,
  V.~Petrov, Phys. Rev. D {\protect \bf 50}, 266 (1994).}%
  \bibAnnoteFile{Stop}{Note2}%
\bibitem{RSTcheck}%
  \BibitemOpen
  \bibfield{author}{%
  \bibinfo {author} {\bibfnamefont{P.~G.}\ \bibnamefont{Tinyakov}},\ }%
  \bibfield{journal}{%
  \bibinfo {journal} {Phys. Lett. B}\ }%
  \textbf{\bibinfo {volume} {284}},\ \bibinfo {pages} {410 } (\bibinfo {year}
  {1992})%
  \bibAnnoteFile{NoStop}{RSTcheck}%
\bibitem{RSTcheck1}%
  \BibitemOpen
  \bibfield{author}{%
  \bibinfo {author} {\bibfnamefont{A.~H.}\ \bibnamefont{Mueller}},\ }%
  \bibfield{journal}{%
  \bibinfo {journal} {Nucl. Phys. B}\ }%
  \textbf{\bibinfo {volume} {401}},\ \bibinfo {pages} {93 } (\bibinfo {year}
  {1993})%
  \bibAnnoteFile{NoStop}{RSTcheck1}%
\bibitem{Bonini}%
  \BibitemOpen
  \bibfield{author}{%
  \bibinfo {author} {\bibfnamefont{G.~F.}\ \bibnamefont{Bonini}}, \bibinfo
  {author} {\bibfnamefont{A.~G.}\ \bibnamefont{Cohen}}, \bibinfo {author}
  {\bibfnamefont{C.}~\bibnamefont{Rebbi}},\ and\ \bibinfo {author}
  {\bibfnamefont{V.~A.}\ \bibnamefont{Rubakov}},\ }%
  \bibfield{journal}{%
  \bibinfo {journal} {Phys. Rev. D}\ }%
  \textbf{\bibinfo {volume} {60}},\ \bibinfo {pages} {076004} (\bibinfo {year}
  {1999})%
  \bibAnnoteFile{NoStop}{Bonini}%
\bibitem{LPS}%
  \BibitemOpen
  \bibfield{author}{%
  \bibinfo {author} {\bibfnamefont{D.~G.}\ \bibnamefont{Levkov}}, \bibinfo
  {author} {\bibfnamefont{A.~G.}\ \bibnamefont{Panin}},\ and\ \bibinfo {author}
  {\bibfnamefont{S.~M.}\ \bibnamefont{Sibiryakov}},\ }%
  \bibfield{journal}{%
  \bibinfo {journal} {J. Phys. A: Math. Theor.}\ }%
  \textbf{\bibinfo {volume} {42}},\ \bibinfo {pages} {205102} (\bibinfo {year}
  {2009})%
  \bibAnnoteFile{NoStop}{LPS}%
\bibitem{Note3}%
  \BibitemOpen
  \bibinfo {note} {The saddle--point method is justified at small $g^2$ when
  the integrand $\protect \qopname \relax o{exp}(iS[\phi ])$ in the path
  integral rapidly oscillates, cf. Eq.~(\ref {eq:2}).}%
  \bibAnnoteFile{Stop}{Note3}%
\bibitem{kink_chaos}%
  \BibitemOpen
  \bibfield{author}{%
  \bibinfo {author} {\bibfnamefont{D.~K.}\ \bibnamefont{Campbell}}, \bibinfo
  {author} {\bibfnamefont{J.~F.}\ \bibnamefont{Schonfeld}},\ and\ \bibinfo
  {author} {\bibfnamefont{C.~A.}\ \bibnamefont{Wingate}},\ }%
  \bibfield{journal}{%
  \bibinfo {journal} {Physica D}\ }%
  \textbf{\bibinfo {volume} {9}},\ \bibinfo {pages} {1 } (\bibinfo {year}
  {1983})%
  \bibAnnoteFile{NoStop}{kink_chaos}%
\bibitem{chaos_semiclassical}%
  \BibitemOpen
  \bibfield{author}{%
  \bibinfo {author} {\bibfnamefont{O.}~\bibnamefont{Bohigas}}, \bibinfo
  {author} {\bibfnamefont{S.}~\bibnamefont{Tomsovic}},\ and\ \bibinfo {author}
  {\bibfnamefont{D.}~\bibnamefont{Ullmo}},\ }%
  \bibfield{journal}{%
  \bibinfo {journal} {Phys. Rep.}\ }%
  \textbf{\bibinfo {volume} {223}},\ \bibinfo {pages} {43 } (\bibinfo {year}
  {1993})%
  \bibAnnoteFile{NoStop}{chaos_semiclassical}%
\bibitem{chaos_semiclassical1}%
  \BibitemOpen
  \bibfield{author}{%
  \bibinfo {author} {\bibfnamefont{A.}~\bibnamefont{Shudo}}\ and\ \bibinfo
  {author} {\bibfnamefont{K.~S.}\ \bibnamefont{Ikeda}},\ }%
  \bibfield{journal}{%
  \bibinfo {journal} {Phys. Rev. Lett.}\ }%
  \textbf{\bibinfo {volume} {74}},\ \bibinfo {pages} {682} (\bibinfo {year}
  {1995})%
  \bibAnnoteFile{NoStop}{chaos_semiclassical1}%
\bibitem{chaos_semiclassical2}%
  \BibitemOpen
  \bibfield{author}{%
  \bibinfo {author} {\bibfnamefont{A.}~\bibnamefont{Shudo}}\ and\ \bibinfo
  {author} {\bibfnamefont{K.~S.}\ \bibnamefont{Ikeda}},\ }%
  \bibfield{journal}{%
  \bibinfo {journal} {Phys. Rev. Lett.}\ }%
  \textbf{\bibinfo {volume} {76}},\ \bibinfo {pages} {4151} (\bibinfo {year}
  {1996})%
  \bibAnnoteFile{NoStop}{chaos_semiclassical2}%
\bibitem{Onishi}%
  \BibitemOpen
  \bibfield{author}{%
  \bibinfo {author} {\bibfnamefont{T.}~\bibnamefont{Onishi}}, \bibinfo {author}
  {\bibfnamefont{A.}~\bibnamefont{Shudo}}, \bibinfo {author}
  {\bibfnamefont{K.~S.}\ \bibnamefont{Ikeda}},\ and\ \bibinfo {author}
  {\bibfnamefont{K.}~\bibnamefont{Takahashi}},\ }%
  \bibfield{journal}{%
  \bibinfo {journal} {Phys. Rev. E}\ }%
  \textbf{\bibinfo {volume} {64}},\ \bibinfo {pages} {025201} (\bibinfo {year}
  {2001})%
  \bibAnnoteFile{NoStop}{Onishi}%
\bibitem{Onishi1}%
  \BibitemOpen
  \bibfield{author}{%
  \bibinfo {author} {\bibfnamefont{T.}~\bibnamefont{Onishi}}, \bibinfo {author}
  {\bibfnamefont{A.}~\bibnamefont{Shudo}}, \bibinfo {author}
  {\bibfnamefont{K.~S.}\ \bibnamefont{Ikeda}},\ and\ \bibinfo {author}
  {\bibfnamefont{K.}~\bibnamefont{Takahashi}},\ }%
  \bibfield{journal}{%
  \bibinfo {journal} {Phys. Rev. E}\ }%
  \textbf{\bibinfo {volume} {68}},\ \bibinfo {pages} {056211} (\bibinfo {year}
  {2003})%
  \bibAnnoteFile{NoStop}{Onishi1}%
\bibitem{Note4}%
  \BibitemOpen
  \bibinfo {note} {We discretize Eqs.~(\ref {eq:11}), (\ref {eq:9}), (\ref
  {eq:8}) by introducing uniform lattice $N_t\times N_x = 5000\times 250$ of
  extent $L_t\times L_x = 15\times 7$. Precision of discretization is kept
  smaller than $0.1\%$; it is controlled by changing the lattice size and
  keeping track of energy conservation. Linear evolution in Eq.~(\ref {eq:12})
  holds with accuracy smaller than $1\%$. Extrapolation $\delta \rho \to 0$
  produces relative errors in $F(E)$ of order $10^{-3}$.}%
  \bibAnnoteFile{Stop}{Note4}%
\bibitem{SU2}%
  \BibitemOpen
  \bibfield{author}{%
  \bibinfo {author} {\bibfnamefont{F.}~\bibnamefont{Bezrukov}}, \bibinfo
  {author} {\bibfnamefont{D.}~\bibnamefont{Levkov}}, \bibinfo {author}
  {\bibfnamefont{C.}~\bibnamefont{Rebbi}}, \bibinfo {author}
  {\bibfnamefont{V.}~\bibnamefont{Rubakov}},\ and\ \bibinfo {author}
  {\bibfnamefont{P.}~\bibnamefont{Tinyakov}},\ }%
  \bibfield{journal}{%
  \bibinfo {journal} {Phys. Rev. D}\ }%
  \textbf{\bibinfo {volume} {68}},\ \bibinfo {pages} {036005} (\bibinfo {year}
  {2003})%
  \bibAnnoteFile{NoStop}{SU2}%
\bibitem{SU21}%
  \BibitemOpen
  \bibfield{author}{%
  \bibinfo {author} {\bibfnamefont{F.}~\bibnamefont{Bezrukov}}, \bibinfo
  {author} {\bibfnamefont{D.}~\bibnamefont{Levkov}}, \bibinfo {author}
  {\bibfnamefont{C.}~\bibnamefont{Rebbi}}, \bibinfo {author}
  {\bibfnamefont{V.}~\bibnamefont{Rubakov}},\ and\ \bibinfo {author}
  {\bibfnamefont{P.}~\bibnamefont{Tinyakov}},\ }%
  \bibfield{journal}{%
  \bibinfo {journal} {Phys. Lett. B}\ }%
  \textbf{\bibinfo {volume} {574}},\ \bibinfo {pages} {75 } (\bibinfo {year}
  {2003})%
  \bibAnnoteFile{NoStop}{SU21}%
\bibitem{Note5}%
  \BibitemOpen
  \bibinfo {note} {One can argue~\cite {RST,Levkov} that $g^2N \cdot \theta \to
  0$ as $g^2 N\to 0$, so we use only the first two terms in Eq.~(\ref
  {eq:8}).}%
  \bibAnnoteFile{Stop}{Note5}%
\bibitem{epsilon}%
  \BibitemOpen
  \bibfield{author}{%
  \bibinfo {author} {\bibfnamefont{F.}~\bibnamefont{Bezrukov}}\ and\ \bibinfo
  {author} {\bibfnamefont{D.}~\bibnamefont{Levkov}},\ }%
  \bibfield{journal}{%
  \bibinfo {journal} {arXiv:quant-ph/0301022~}}%
   (\bibinfo {year} {2003})%
  \bibAnnoteFile{NoStop}{epsilon}%
\bibitem{epsilon1}%
  \BibitemOpen
  \bibfield{author}{%
  \bibinfo {author} {\bibfnamefont{F.}~\bibnamefont{Bezrukov}}\ and\ \bibinfo
  {author} {\bibfnamefont{D.}~\bibnamefont{Levkov}},\ }%
  \bibfield{journal}{%
  \bibinfo {journal} {JETP}\ }%
  \textbf{\bibinfo {volume} {98}},\ \bibinfo {pages} {820} (\bibinfo {year}
  {2004})%
  \bibAnnoteFile{NoStop}{epsilon1}%
\bibitem{epsilon2}%
  \BibitemOpen
  \bibfield{author}{%
  \bibinfo {author} {\bibfnamefont{D.~G.}\ \bibnamefont{Levkov}}, \bibinfo
  {author} {\bibfnamefont{A.~G.}\ \bibnamefont{Panin}},\ and\ \bibinfo {author}
  {\bibfnamefont{S.~M.}\ \bibnamefont{Sibiryakov}},\ }%
  \bibfield{journal}{%
  \bibinfo {journal} {Phys. Rev. Lett.}\ }%
  \textbf{\bibinfo {volume} {99}},\ \bibinfo {pages} {170407} (\bibinfo {year}
  {2007})%
  \bibAnnoteFile{NoStop}{epsilon2}%
\bibitem{Note6}%
  \BibitemOpen
  \bibinfo {note} {The critical bubble decays in the right part of Fig.~\ref
  {solutions}e due to regularization $\epsilon >0$.}%
  \bibAnnoteFile{Stop}{Note6}%
\bibitem{new_mechanism}%
  \BibitemOpen
  \bibfield{author}{%
  \bibinfo {author} {\bibfnamefont{K.}~\bibnamefont{Takahashi}}\ and\ \bibinfo
  {author} {\bibfnamefont{K.~S.}\ \bibnamefont{Ikeda}},\ }%
  \bibfield{journal}{%
  \bibinfo {journal} {J. Phys. A: Math. Gen.}\ }%
  \textbf{\bibinfo {volume} {36}},\ \bibinfo {pages} {7953} (\bibinfo {year}
  {2003})%
  \bibAnnoteFile{NoStop}{new_mechanism}%
\bibitem{new_mechanism1}%
  \BibitemOpen
  \bibfield{author}{%
  \bibinfo {author} {\bibfnamefont{K.}~\bibnamefont{Takahashi}}\ and\ \bibinfo
  {author} {\bibfnamefont{K.~S.}\ \bibnamefont{Ikeda}},\ }%
  \bibfield{journal}{%
  \bibinfo {journal} {EPL}\ }%
  \textbf{\bibinfo {volume} {71}},\ \bibinfo {pages} {193} (\bibinfo {year}
  {2005})%
  \bibAnnoteFile{NoStop}{new_mechanism1}%
\bibitem{Levkov}%
  \BibitemOpen
  \bibfield{author}{%
  \bibinfo {author} {\bibfnamefont{D.~G.}\ \bibnamefont{Levkov}}\ and\ \bibinfo
  {author} {\bibfnamefont{S.~M.}\ \bibnamefont{Sibiryakov}},\ }%
  \bibfield{journal}{%
  \bibinfo {journal} {Phys. Rev. D}\ }%
  \textbf{\bibinfo {volume} {71}},\ \bibinfo {pages} {025001} (\bibinfo {year}
  {2005})%
  \bibAnnoteFile{NoStop}{Levkov}%
\bibitem{Levkov1}%
  \BibitemOpen
  \bibfield{author}{%
  \bibinfo {author} {\bibfnamefont{D.~G.}\ \bibnamefont{Levkov}}\ and\ \bibinfo
  {author} {\bibfnamefont{S.~M.}\ \bibnamefont{Sibiryakov}},\ }%
  \bibfield{journal}{%
  \bibinfo {journal} {JETP Lett.}\ }%
  \textbf{\bibinfo {volume} {81}},\ \bibinfo {pages} {53} (\bibinfo {year}
  {2005})%
  \bibAnnoteFile{NoStop}{Levkov1}%
\bibitem{Demidov}%
  \BibitemOpen
  \bibfield{author}{%
  \bibinfo {author} {\bibfnamefont{S.~V.}\ \bibnamefont{Demidov}}\ and\
  \bibinfo {author} {\bibfnamefont{D.~G.}\ \bibnamefont{Levkov}},\ }%
  \bibfield{journal}{%
  \bibinfo {journal} {JHEP}\ }%
  \textbf{\bibinfo {volume} {1106}},\ \bibinfo {pages} {016} (\bibinfo {year}
  {2011})%
  \bibAnnoteFile{NoStop}{Demidov}%
\bibitem{Rebbi}%
  \BibitemOpen
  \bibfield{author}{%
  \bibinfo {author} {\bibfnamefont{C.}~\bibnamefont{Rebbi}}\ and\ \bibinfo
  {author} {\bibfnamefont{R.}~\bibnamefont{Singleton}},\ }%
  \bibfield{journal}{%
  \bibinfo {journal} {Phys. Rev. D}\ }%
  \textbf{\bibinfo {volume} {54}},\ \bibinfo {pages} {1020} (\bibinfo {year}
  {1996})%
  \bibAnnoteFile{NoStop}{Rebbi}%
\bibitem{Shifman}%
  \BibitemOpen
  \bibfield{author}{%
  \bibinfo {author} {\bibfnamefont{M.}~\bibnamefont{Maggiore}}\ and\ \bibinfo
  {author} {\bibfnamefont{M.}~\bibnamefont{Shifman}},\ }%
  \bibfield{journal}{%
  \bibinfo {journal} {Nucl. Phys. B}\ }%
  \textbf{\bibinfo {volume} {371}},\ \bibinfo {pages} {177 } (\bibinfo {year}
  {1992})%
  \bibAnnoteFile{NoStop}{Shifman}%
\bibitem{Manton}%
  \BibitemOpen
  \bibfield{author}{%
  \bibinfo {author} {\bibfnamefont{N.~S.}\ \bibnamefont{Manton}},\ }%
  \bibfield{journal}{%
  \bibinfo {journal} {Nucl. Phys. B}\ }%
  \textbf{\bibinfo {volume} {126}},\ \bibinfo {pages} {525} (\bibinfo {year}
  {1977})%
  \bibAnnoteFile{NoStop}{Manton}%
\bibitem{Monopoles}%
  \BibitemOpen
  \bibfield{author}{%
  \bibinfo {author} {\bibfnamefont{I.~K.}\ \bibnamefont{Affleck}}\ and\
  \bibinfo {author} {\bibfnamefont{N.~S.}\ \bibnamefont{Manton}},\ }%
  \bibfield{journal}{%
  \bibinfo {journal} {Nucl. Phys. B}\ }%
  \textbf{\bibinfo {volume} {194}},\ \bibinfo {pages} {38 } (\bibinfo {year}
  {1982})%
  \bibAnnoteFile{NoStop}{Monopoles}%
\bibitem{schwinger_induced}%
  \BibitemOpen
  \bibfield{author}{%
  \bibinfo {author} {\bibfnamefont{A.~K.}\ \bibnamefont{Monin}},\ }%
  \bibfield{journal}{%
  \bibinfo {journal} {JHEP}\ }%
  \textbf{\bibinfo {volume} {0510}},\ \bibinfo {pages} {109} (\bibinfo {year}
  {2005})%
  \bibAnnoteFile{NoStop}{schwinger_induced}%
\end{thebibliography}%
\end{document}